\documentclass{moriond}

\def\Journal#1#2#3#4{{#1} {\bf #2}, #3 (#4)}

\def\be{\begin{equation}}
\def\ee{\end{equation}}
\def\bea{\begin{eqnarray}}
\def\eea{\end{eqnarray}}

\newcommand{\Photo}{}

\begin{document}
\vspace*{4cm}
\title{Component Separation method for CMB using Convolutional Neural Networks}

\author{ A. Quintana\textsuperscript{1,2,3}, B. Ruiz-Granados\textsuperscript{4}, P. Ruiz-Lapuente\textsuperscript{1,2}}

\address{\textsuperscript{1}Instituto de Física Fundamental (IFF-CSIC), Madrid, Spain,\\
\textsuperscript{2}Institut de Ciències del Cosmos (UB-IEEC), Barcelona, Spain\\
\textsuperscript{3}Universidad Internacional de Valencia, Valencia, Spain\\
\textsuperscript{4}Universidad de Córdoba, Córdoba, Spain}

\maketitle\abstracts{
The aim of this project is to recover the CMB anisotropies maps in temperature and polarized intensity by means of a deep convolutional neural network (CNN) which, after appropiate training, can remove the foregrounds from Planck and QUIJOTE data. The results are then compared with those obtained by \texttt{COMMANDER}, based on Bayesian parametric component separation. The CNN successfully recovered the CMB signal for both All Sky and Partial Sky maps showing frequency dependant results, being optimum for central frequencies where there is less contamination by foregrounds emissions such as galactic synchrotron and thermal dust emissions. Recovered maps in temperature are consistent with those obtained by Planck Collaboration, while polarized intensity has been recovered as a new observable. The polarized intensity maps recovered from QUIJOTE experiment are novel and of potential interest to the scientific community for the detection of primordial gravitational waves. The way forward will be to recover the maps at higher NSIDE and make them available to the scientific community.}

\section{Introduction}
Previous works \cite{wang} have successfully demonstrated the use of Deep Learning techniques, such as Convolutional Neural Networks (CNN), for the extraction of CMB temperature maps, in particular with \texttt{U-Net} architecture \cite{ronneberger}. In this work a \texttt{U-Net} CNN is trained from simulated observational maps, with the aim of recovering the CMB signal in temperature and polarized intensity from Planck \footnote[1]{https://www.cosmos.esa.int/web/planck} and QUIJOTE \footnote[2]{https://www.iac.es/es/proyectos/experimento-quijote-cmb} observational maps. In order to validate the methodology, the recovered CMB maps are compared with those obtained by Planck Collaboration with \texttt{COMMANDER} as a different Component Separation Method.

\section{Methodology and Results}
A CNN with \texttt{U-Net} architecture is developed and trained in order to be able to recover the CMB signal. To train the CNN, 803 mock maps are simulated at NSIDE 64 at different frequency bands, both in temperature and polarized intensity for Planck and QUIJOTE maps, meaning that the CNN has to be trained four times. A different dataset of 209 mock maps are also generated in order to validate the CNN, and finally 11 mock maps are generated for testing the CNN. For each frequency, a different realization of the clean CMB signal generated with \texttt{PySM} is provided to the CNN. Once the CNN has been proved to be properly trained with the validation and test dataset, it is directly applied to resolve the Planck and QUIJOTE maps.

\begin{figure}
\centering
\begin{minipage}{0.41\linewidth}
\centerline{\includegraphics[width=1\linewidth]{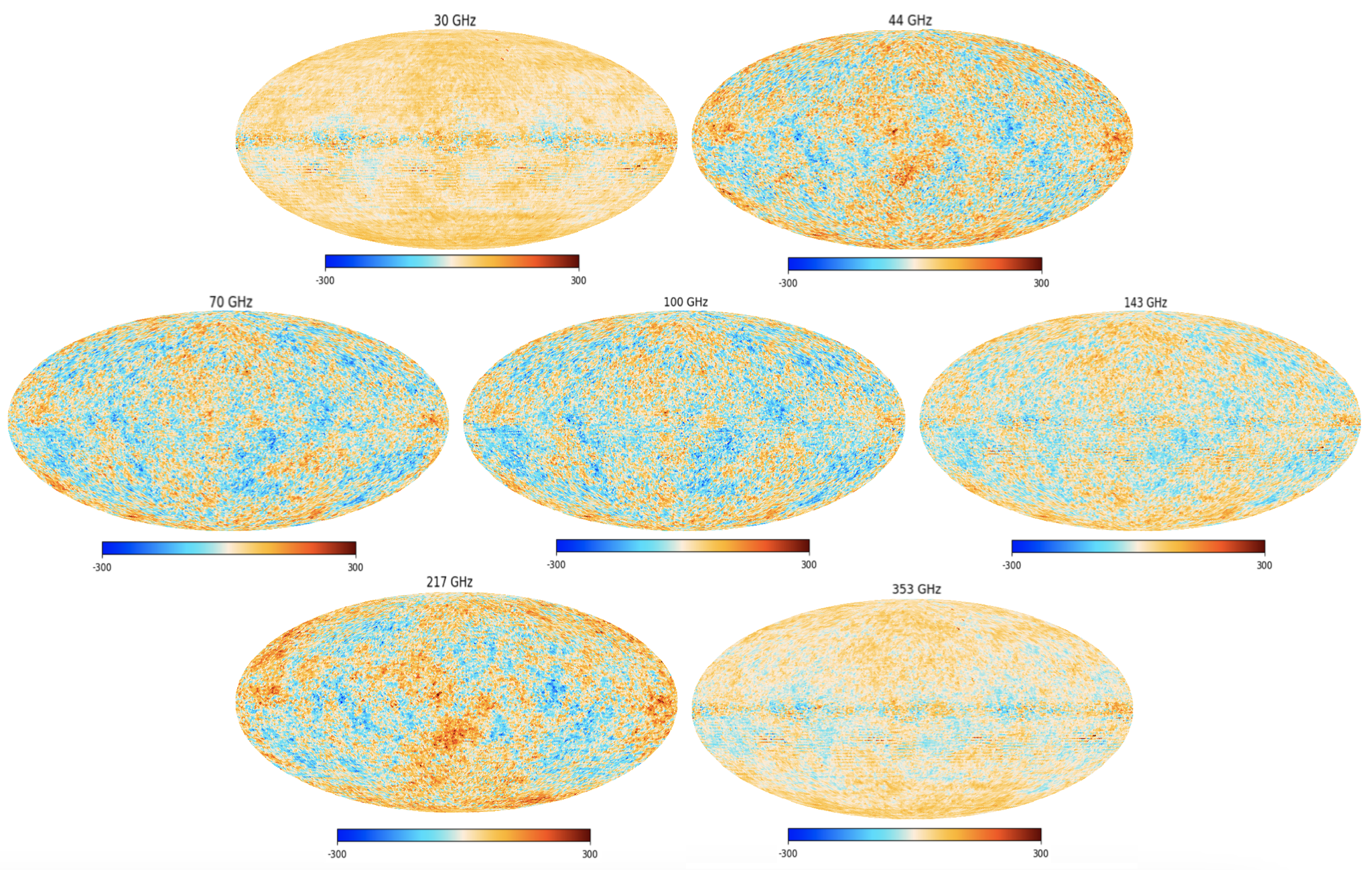}}
\end{minipage}
\begin{minipage}{0.33\linewidth}
\centerline{\includegraphics[width=1\linewidth]{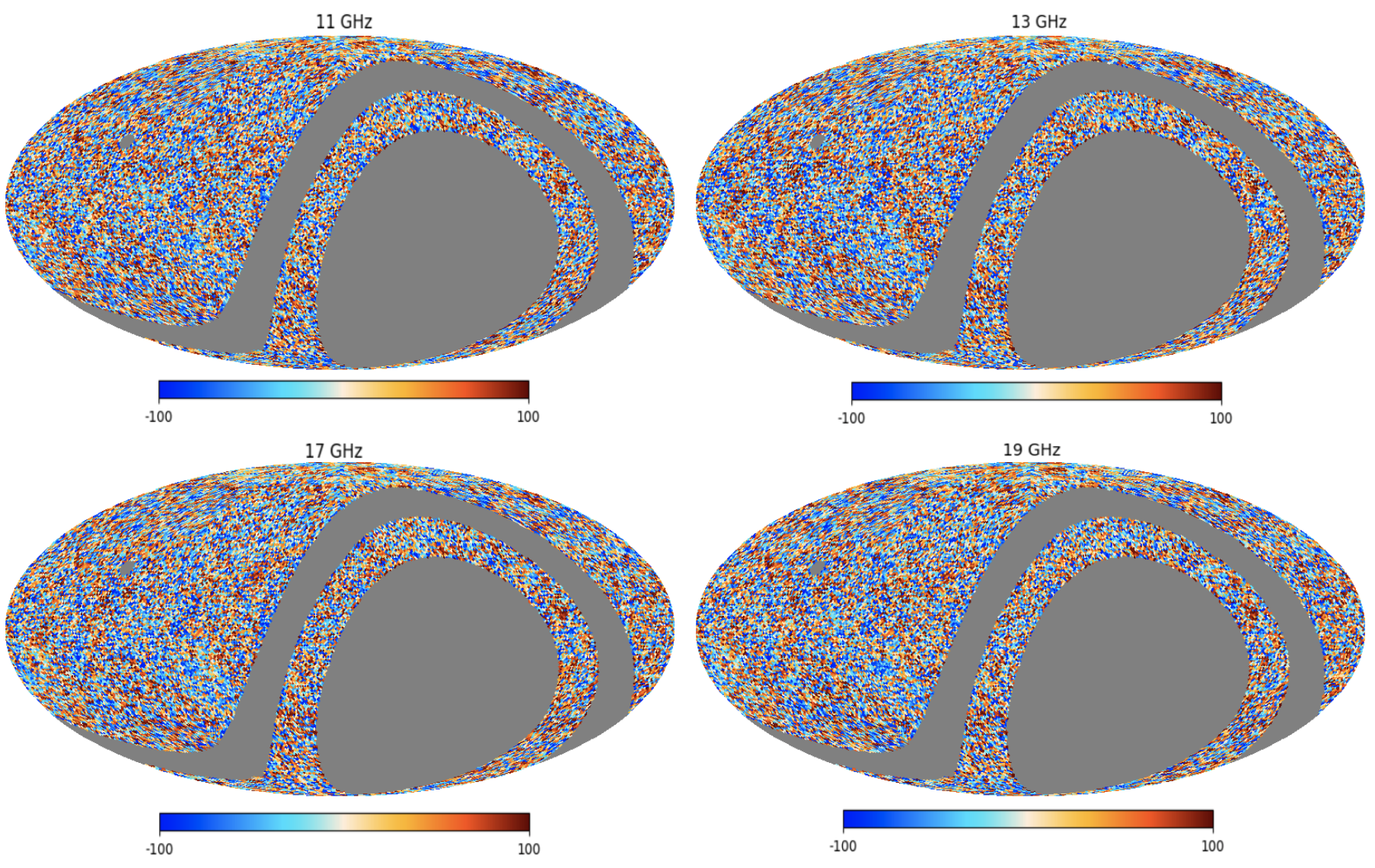}}
\end{minipage}
\caption[]{Left: Recovered CMB maps in temperature at frequencies 30, 44, 70, 100, 143, 217 and 353 GHz from Planck. Right: Recovered CMB maps in polarized intensity at frequencies 11, 13, 17 and 19 GHz from QUIJOTE.}
\label{planck_maps}
\end{figure}

\begin{figure}
    \centering
    \includegraphics[scale=0.18]{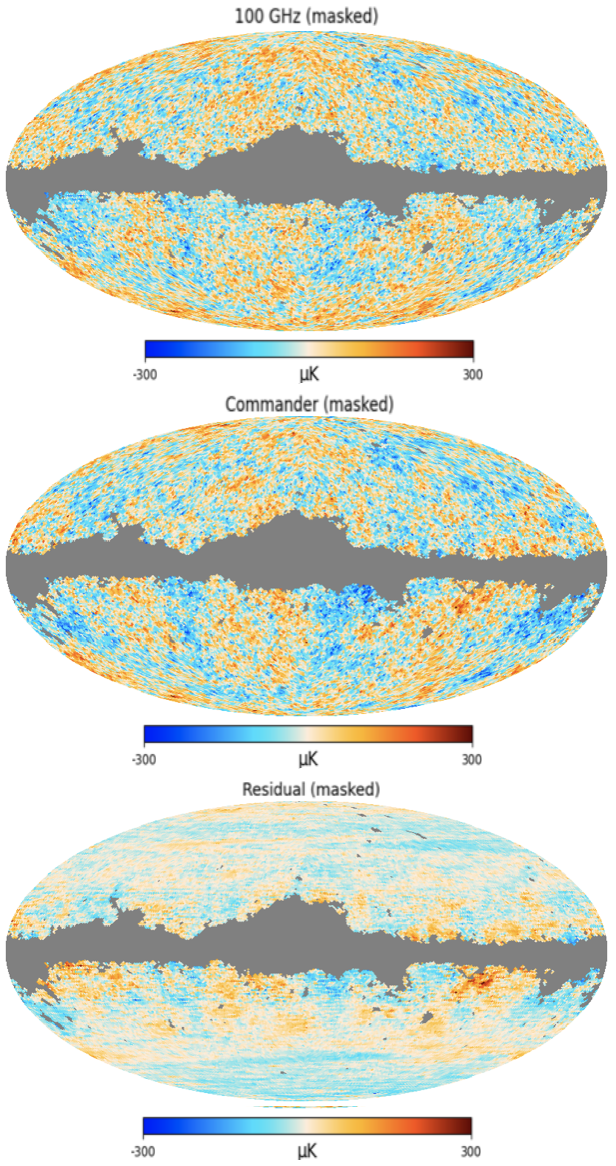}
    \includegraphics[scale=0.15]{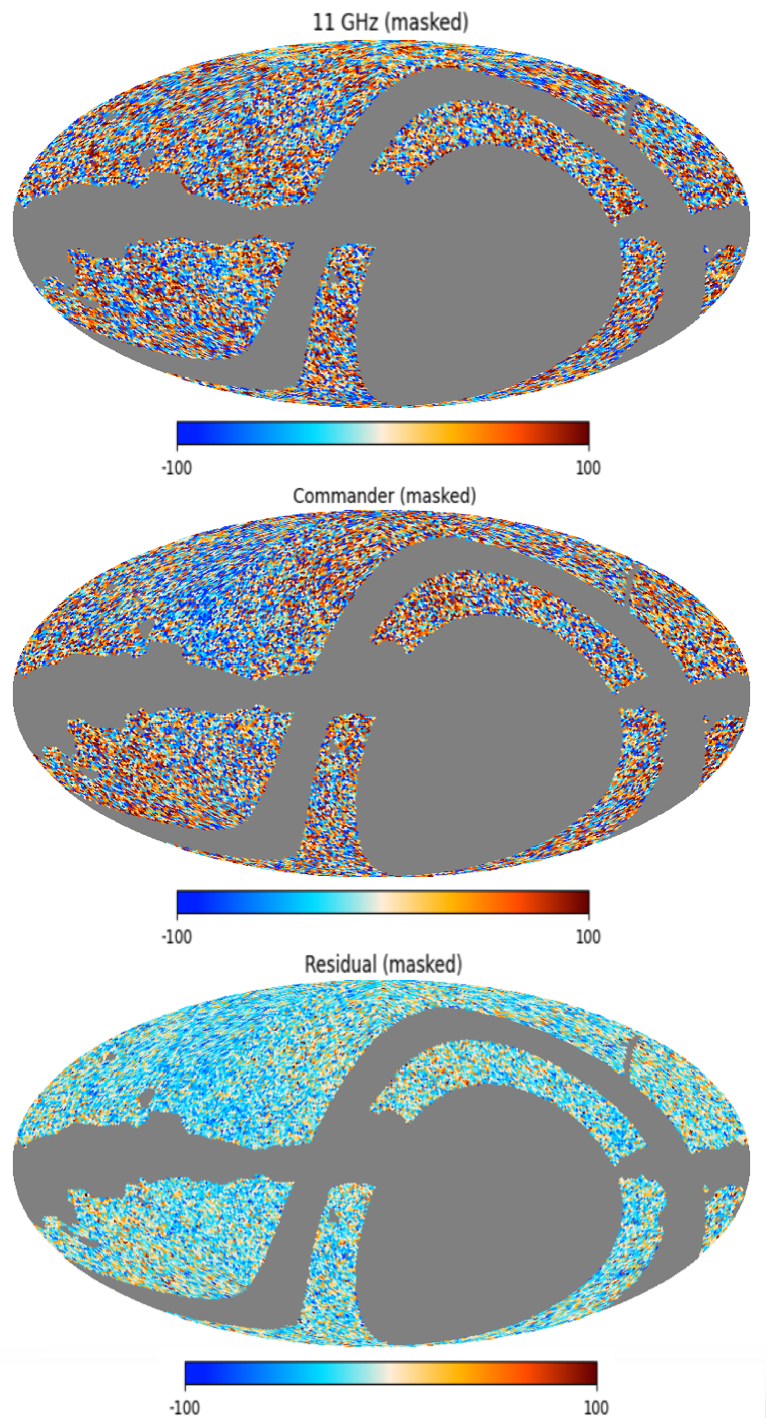}   
    \caption[]{Comparison between CMB signal in temperature recovered from Planck (left) and in polarized intensity from QUIJOTE (right). CMB recovered by the CNN (up), by \texttt{COMMANDER} (middle) and residuals (down) are shown.}
    \label{comparativa_commander}
\end{figure}

In Figure \ref{planck_maps}, it is observed that below 44 GHz and above 217 GHZ the CNN struggles to recover the CMB signal in temperature since synchrotron and thermal dust foregrounds emissions, respectively, are dominant. In Figure \ref{comparativa_commander}, on the left side it is seen that recovered map in temperature is consistent with the CMB obtained by Planck. On the right side, while deviations from results in polarized intensity with respect to Planck are apparent by eye, it can be proved that the map recovered by the CNN is significantly more Gaussian than Planck, which could be a hint of a better CMB signal; deep statistical analysis is yet to be performed though.

\section{Conclusions}
CMB anisotropies maps have been successfully recovered by a CNN for both All Sky and Partial Sky maps. Recovered maps are frequency dependant, with optimum results for central frequencies. It is the first time that polarized intensity defined as a scalar $IP=\sqrt{Q^2+U^2}$ is recovered as an observable. While all simulations have been run at NSIDE 64, the next step will be to repeat the procedure at NSIDE 512.

\end{document}